\documentclass[aps,prc,superscriptaddress,showpacs,twocolumn,floatfix]{revtex4}

\usepackage{graphicx}

\def\gtorder{\mathrel{\raise.3ex\hbox{$>$}\mkern-14mu
 \lower0.6ex\hbox{$\sim$}}}
\def\ltorder{\mathrel{\raise.3ex\hbox{$<$}\mkern-14mu
 \lower0.6ex\hbox{$\sim$}}}
\def\gep{G_{Ep}}
\def\gmp{G_{Mp}}
\def\gegm{G_{Ep} / G_{Mp}}
\def\mugegm{\mu_p G_{Ep} / G_{Mp}}

\def\etal{\emph{et al.}}

\begin{document}

\title{Reexamination of phenomenological two-photon exchange corrections to
the proton form factors and $e^{\pm}p$ scattering}

\author{I.~A.~Qattan}
\affiliation{Khalifa University of Science, Technology and Research, Department
of Physics, P.O. Box 573, Sharjah, U.A.E}

\author{A.~Alsaad}
\affiliation{Jordan University of Science and Technology, Department of
Physical Sciences, P.O. Box 3030, Irbid 22110, Jordan}

\author{J.~Arrington}
\affiliation{Physics Division, Argonne National Laboratory, Argonne, Illinois,
60439, USA}

\date{\today} 


\begin{abstract}

We extract the two-photon exchange (TPE) contributions to electron--proton
elastic scattering using two parametrizations and compare the results to
different phenomenological extractions and direct calculations of the TPE
effects.  We find that many of the extractions give similar results, and
highlight the common assumptions and the impact of not including these
assumptions.  We provide a simple parametrization of the TPE contribution
to the unpolarized cross section, along with an estimate of the fit
uncertainties and the uncertainties associated with the model dependence of
the extractions. We look at the contributions as extracted from various e--p
elastic scattering observables, and make predictions for ratio of
positron-proton to electron-proton elastic scattering cross sections.

\end{abstract}

\pacs{25.30.Bf, 13.40.Gp, 14.20.Dh}

\maketitle

\section{Introduction}

Electron scattering is a powerful technique used to reveal the underlying
structure of the nucleon. This is because the electron is a point-like,
structureless particle.  This makes it an ideal probe of the target, as the
cross section represents only known couplings and the (unknown) structure of
the target.  In elastic electron scattering, the incident electron scatters
from a nucleon through the exchange of a virtual photon $\gamma^{*}$,
and the structure of the target nucleon appears through two
Sachs~\cite{sachs62} electromagnetic form factors, $G_{Ep}$ and $G_{Mp}$,
which measure the deviation of the scattering from that for a point-like
spin-1/2 target. As $Q^2$, the four-momentum squared of the virtual photon,
increases the scattering becomes more sensitive to the small scale internal
structure of the target.

Utilizing electron scattering, there are two methods used to extract the
proton's form factors. The first is the Rosenbluth or Longitudinal-Transverse
(LT) separation method~\cite{rosenbluth50}, which uses measurements of the
unpolarized cross section, and the second is the polarization transfer or
polarized target (PT) method~\cite{dombey69}, which requires measurement of
the spin-dependent cross section. In the Rosenbluth separation method, the
reduced cross section $\sigma_{R}$ is defined in the one-photon exchange (OPE)
approximation as
\begin{equation} \label{eq:reduced2}
\sigma_{R} = G_{Mp}^{2}(Q^2) + \frac{\varepsilon}{\tau} G_{Ep}^{2}(Q^2) ,
\end{equation}
where $\tau = Q^2/4M_{p}^2$, $M_{p}$ is the mass of the proton, and
$\varepsilon$ is the virtual photon longitudinal polarization parameter,
defined as $\varepsilon^{-1} = \big[ 1 + 2 (1+\tau)
\tan^{2}({\frac{\theta_{e}}{2}})\big]$, where $\theta_{e}$ is the scattering
angle of the electron. For a fixed $Q^{2}$ value, the reduced cross section
$\sigma_{R}$ is measured at several $\varepsilon$ points, and a linear fit of
$\sigma_{R}$ to $\varepsilon$ gives $G_{Mp}^2(Q^2)$ as the intercept and
$G_{Ep}^2(Q^2)/\tau$ as the slope.

In the recoil polarization method, a beam of longitudinally polarized
electrons scatters elastically from unpolarized proton target. The electrons
transfer their polarization to the unpolarized protons. By simultaneously
measuring the transverse, $P_{t}$, and longitudinal, $P_{l}$, polarization
components of the recoil proton, one can determine the ratio
$\mu_{p}G_{Ep}/G_{Mp}$ in the OPE~\cite{dombey69,akhiezer74,arnold81}:
\begin{equation} \label{eq:ratio}
\frac{G_{Ep}}{G_{Mp}} = -\frac{P_{t}}{P_{l}} \frac{(E+E')}{2M_{p}} \tan({\frac{\theta_{e}}{2}}),
\end{equation}
where $E$ and $E'$ are the initial and final energy of the incident electron,
respectively.  The ratio can be extracted in a similar fashion using polarized
beams and targets by measuring the asymmetry for two different spin
directions~\cite{arrington07a,perdrisat07}.

The two methods yield significantly and strikingly different results in the
region $Q^2 \geq$ 1.0 (GeV/c)$^2$, with values of $\mu_{p}G_{Ep}/G_{Mp}$
differing almost by a factor of three at high $Q^2$. In the LT separation
method, the ratio shows approximate form factor scaling, $\mu_{p}G_{Ep}/G_{Mp}
\approx 1$, albeit with large uncertainties at high $Q^2$ values. On the other
hand, the recoil polarization method yields a ratio that decreases with
increasing $Q^2$. The ratio is well parametrized~\cite{gayou02} as
\begin{equation} \label{eq:ratio2}
\frac{\mu_{p}G_{Ep}}{G_{Mp}} = \mu_{p}R = 1 - 0.13(Q^2-0.04).
\end{equation}

To reconcile the ratios, several studies suggested that missing higher order
radiative corrections to the electron-proton elastic scattering cross sections
can explain the discrepancy. In particular, studies focused on the role of the
two-photon exchange (TPE) effect in resolving the discrepancy. The TPE effect
was studied extensively both theoretically~\cite{blunden03, blunden05a,
kondratyuk05, chen04, borisyuk06a, borisyuk06b} and phenomenologically
\cite{guichon03, gustafsson05a, tvaskis06, borisyuk06c, chen07, arrington05,
guttmann2011, qattan2011, borisyuk2011}. An extensive review of the role of the
TPE effect in electron-proton scattering can be found in~\cite{arrington11b,
carlson07}.

\section{Two-photon-exchange and the $e^{\pm}p$ cross section scattering
ratio}  \label{visit_interference}

The interference of the OPE and TPE amplitudes represents the leading TPE
correction to the electron-proton elastic scattering cross section. However,
the recoil polarization data were confirmed experimentally to be essentially
independent of $\varepsilon$~\cite{meziane2011}. We account for the TPE
contribution to $\sigma_{R}$ simply by adding the real function
$F(Q^2,\varepsilon)$ to the Born reduced cross section:
\begin{equation} \label{eq:reduced3}
\sigma_{R} = G_{Mp}^{2}\Big(1 + \frac{\varepsilon}{\tau} R^2\Big) + F(Q^2,\varepsilon),
\end{equation}
where $R = G_{Ep}/G_{Mp}$ is the recoil polarization ratio Eq.
(\ref{eq:ratio2}). The TPE correction  changes sign for electron-proton and
positron-proton scattering, so the ratio $R_{e^{+} e^{-}}(Q^2,\varepsilon)$
defined as
\begin{equation} \label{eq:ratiopostelect}
R_{e^{+} e^{-}}(Q^2,\varepsilon) = \frac{\sigma(e^{+}p~\rightarrow e^{+}p)}
{\sigma(e^{-}p~\rightarrow e^{-}p)},  
\end{equation}  
allows for direct measurements of the TPE contribution to the cross section.
The ratio $R_{e^{+} e^{-}}$
can be expressed as $R_{e^{+} e^{-}} \approx 1+4\Re(A_{2\gamma})/A_{1\gamma}$,
with $A_{1\gamma}$ and $A_{2\gamma}$ being the OPE and TPE
amplitudes~\cite{mar68}, respectively. Here $\Re$ stands for the real part. On
the other hand, the modification to the electron cross section is 
$1-2\Re(A_{2\gamma})/A_{1\gamma}$. Clearly, any change in the electron cross
section will have almost twice the change in the ratio $R_{e^{+} e^{-}}$ but
with the opposite sign.


Recently, several theoretical studies estimated the function
$F(Q^2,\varepsilon)$~\cite{blunden03, blunden05a, kondratyuk05, chen04,
borisyuk06a,borisyuk06b}. Experimentally, studies focused on searching for
nonlinearities in $\sigma_{R}$~\cite{qattanphd, tvaskis06, gustafsson05a,
chen07} and phenomenologically by reanalyzing the experimental data using a
proposed parametrization of the TPE contribution~\cite{guichon03,
gustafsson05a, borisyuk06c, chen07, arrington05, guttmann2011, qattan2011,
borisyuk2011}.  We present a summary of the results which we will examine in
detail in this work.

Guichon and Vanderhaeghen~\cite{guichon03} expressed the hadronic vertex
function in terms of three independent complex amplitudes (generalized form
factors) which depend on both $Q^2$ and $\varepsilon$:
$\tilde{G}_{Ep}(\varepsilon,Q^2)$, $\tilde{G}_{Mp}(\varepsilon,Q^2)$, and
$\tilde{F_{3}}(\varepsilon,Q^2)$. These generalized form factors can be broken
into the usual Born (OPE) and the TPE contributions as:
$\tilde{G}_{Ep,Mp}(\varepsilon,Q^2)= G_{Ep,Mp}(Q^2) + \Delta
G_{Ep,Mp}(\varepsilon,Q^2)$ with $Y_{2\gamma}(\nu,Q^2)$ defined as $\Re \Big(
\frac{\nu \tilde{F_{3}}} {M^2_{p}|G_{Mp}|} \Big)$, where
$\nu =M^2_{p}\sqrt{(1+\varepsilon)/(1-\varepsilon)}
\sqrt{\tau(1+\tau)}$. The reduced cross section is expressed in terms of these
amplitudes as
\begin{equation} \label{eq:Guichon1}
\sigma_{R} = |\tilde{G}_{Mp}|^2 \Bigg[ 1 + \frac{\varepsilon}{\tau} \frac{|\tilde{G}_{Ep}|^2}{|\tilde{G}_{Mp}|^2} + 2\varepsilon \Bigg(1 + \frac{|\tilde{G}_{Ep}|}{\tau |\tilde{G}_{Mp}|}\Bigg) Y_{2\gamma} \Bigg].
\end{equation}
They demonstrated that small TPE contributions could significantly modify
the extraction of the form factor ratio from the Rosenbluth separation while
having relatively little impact on the polarization transfer data.  They
also provided an estimate of the TPE amplitudes, based on the assumption that
the entire contribution comes from the $Y_{2\gamma}(\nu,Q^2)$ term,
parametrized so as to yield a correction to $\sigma_R$ that is proportional
to $\varepsilon$.  This yields a small reduction in $\gep$ as extracted
from the polarization measurement, and a significant reduction in the
Rosenbluth extraction of $\gep$, such that it becomes consistent with the
corrected polarization results.  In this work, $\gmp$ is unaffected by
the TPE contributions.

Based on the framework of~\cite{guichon03}, Arrington~\cite{arrington05}
performed a global analysis where he extracted the TPE amplitudes $\Delta
G_{Ep,Mp}$ and $Y_{2\gamma}$.  He assumed that the amplitudes were
$\varepsilon$-independent and took $\Delta G_{Ep}=0$. Values for
$Y_{2\gamma}(Q^2)$ were extracted from the difference between polarization and
Rosenbluth measurements, taking into account the uncertainties in both data
sets. Based on the high-$\varepsilon$ constraints from the comparison of
positron and electron scattering~\cite{arrington04b} the amplitude $\Delta
G_{Mp}$ was determined by requiring that its contribution to $\sigma_{R}$ at
$\varepsilon=1$ cancelled the contribution of $Y_{2\gamma}$.  The extracted
TPE amplitudes and their estimated uncertainties are then parametrized as a
function of $Q^2$, and used to apply TPE corrections to the form factors
obtained from a global Rosenbluth analysis~\cite{arrington04a} and the new
recoil polarization data.  Throughout this text, we will refer to the fit to
the uncorrected Rosenbluth form factors obtained in Ref.~\cite{arrington04a}
as the Arrington $\sigma_{R}$ Fit, and to those corrected using the extracted
TPE amplitudes as the Arrington $Y_{2\gamma}$ Fit~\cite{arrington05}.

In both of these analysis, there is not enough information to directly
determine the amplitudes, so assumptions have to be made about the relative
importance and the $\varepsilon$-dependence of the three TPE amplitudes.
Common to these analyses are the assumption that the correction is close to
linear in $\varepsilon$, as no non-linearities have been
observed~\cite{qattanphd, tvaskis06, gustafsson05a, chen07}.  If one also
neglects the TPE correction to the polarization data, which is significantly
smaller at high $Q^2$, then it is not necessary to work in terms of the
polarization amplitude; one can simply parametrize the TPE contributions
to the reduced cross section, taking a linear (or nearly linear) $\varepsilon$
dependence.

Alberico~\etal~\cite{alberico09a} performed a global extraction of the
proton form factors and TPE contributions, based on one of the
parametrizations of the TPE contributions from Chen~\emph{et
al}.~\cite{chen07}:
\begin{equation} \label{eq:chen1}
\sigma_{R} = G_{Mp}^{2}\Big(1 + \frac{\varepsilon}{\tau} R^2\Big) + A(Q^2)y+ B(Q^2)y^3,
\end{equation}
where $y=\sqrt{(1-\varepsilon)/(1+\varepsilon)}$, $A(Q^2)$=$\alpha
G^{2}_D(Q^2)$, $B(Q^2)$=$\beta G^{2}_{D}(Q^2)$ and $G_D(Q^2)$ is the dipole
parameterization: $G_D(Q^2)=[1+Q^2/(0.71(\mbox{GeV/c})^2)]^{-2}$.
Note that Chen~\etal performed such an extraction,
but analyzed only the data of Ref.~\cite{andivahis94}.  Alberico~\etal
extract the form factors and the parameters $\alpha$ and $\beta$ in two
different ways.  In the first fit, referred to throughout this text as ABGG
Fit 1, the polarization transfer data was used to fix the ratio
$R(Q^2)=\mu_{p}G_{Ep}/G_{Mp}=1.022-0.13 Q^2$ and $\gmp$ and the TPE
contribution from Eq.~(\ref{eq:chen1}) were determined in a global fit to
world's cross section data. $G_{Mp}$ was parametrized using the functional
form from Kelly~\cite{kelly04}:
\begin{equation} \label{eq:kellygm}
\frac{G_{Mp}(Q^2)}{\mu_{p}} = \frac{1+ \sum_{k=1}^{n} a_{p,k}^{M} \tau^{k}}{1+ \sum_{k=1}^{n+2} b_{p,k}^{M} \tau^{k}}
\end{equation}
with $n=1$, yielding four parameters for $G_{Mp}$ and the two TPE parameters
$\alpha$ and $\beta$. In their second fit (ABGG Fit 2), the cross section and
polarization transfer data were fit simultaneously, using 4 parameter fits to
$\gep$ and $\gmp$ along with the two TPE parameters.

Qattan and Alsaad~\cite{qattan2011} proposed a different empirical
parametrization for the function $F(Q^2,\varepsilon)$. The function
$F(Q^2,\varepsilon)$ was double Taylor series expanded as a polynomial of
order $n$ keeping only terms linear in $\varepsilon$ but without constraining
the TPE amplitudes by enforcing the Regge limit. In their final result,
labeled fit III, the reduced cross section $\sigma_{R}$ was parametrized as
\begin{equation} \label{eq:qattanfit2}
\sigma_{R} = G_{Mp}^{2}\Big(1 + \frac{\varepsilon}{\tau}R^2\Big) + \varepsilon f(Q^2) ,
\end{equation}
where for the extraction from a fixed-$Q^2$ data set, $f(Q^2)$ is just a
constant. Throughout this text, Eq. (\ref{eq:qattanfit2}) will be referred to
as the QA parametrization.

As noted in some previous extractions~\cite{arrington04b, arrington05,
borisyuk2011, guttmann2011}, small angle comparisons of positron to electron
scattering set tight limits on the TPE effects and TPE
calculations~\cite{blunden03, chen07, carlson07, arrington11b}
show that the contribution should be zero at $\varepsilon=1$.  Therefore, we
perform an updated version of the fit from Ref.~\cite{qattan2011} using a
parametrization which maintains the linear correction in $\varepsilon$ but
yields $F(Q^2,\varepsilon)=0$ at $\varepsilon=1$. Based on the parametrization
from Borisyuk and Kobushkin \cite{borisyuk2011}, we take the following form
\begin{equation} \label{eq:Kobushkin3}
\sigma_{R} = G_{Mp}^{2}\Big( 1 + \frac{\varepsilon}{\tau} R^2\Big) + 2 a(1-\varepsilon)G_{Mp}^{2} ,
\end{equation} 
for our updated extraction. Throughout this text, Eq.~(\ref{eq:Kobushkin3})
will be referred to as the BK parametrization. Because the recoil polarization
ratio $G_{Ep}/G_{Mp}$ was experimentally confirmed to be independent of
$\varepsilon$~\cite{meziane2011}, the ratio $R=G_{Ep}/G_{Mp}$ was fixed to be
that of the recoil polarization ratio or Eq.~(\ref{eq:ratio2}), as in the
analysis of Ref.~\cite{qattan2011}

\section{Results and discussion} \label{results_discussion}
In this article we do the following: 

(1) We extend the analysis of Ref.~\cite{qattan2011}, extracting the TPE
contributions to $\sigma_R$ using both the QA and BK parametrizations for the
TPE amplitude.  We also include additional data sets~\cite{bartel73, litt70,
berger71}, that were not included in the original analysis.  For the data
from Qattan~\etal~\cite{qattan05} we use the published results, including
the $\varepsilon$-dependent systematic uncertainties, yielding somewhat 
modified results than in the previous analysis~\cite{qattan2011}.

(2) We take the proton form factors and TPE amplitudes extracted using the QA
and BK parametrizations and compare them to those obtained using the
extracted TPE corrections Refs.~\cite{arrington04a, alberico09a, arrington05}
and the TPE calculation of Ref.~\cite{arrington07}. We also use our results
and the previous parametrizations to determine the ratio $R_{e^{+} e^{-}}$. 
We use these results to examine the $Q^2$ dependence of the extracted TPE
corrections, the consistency of the different extractions, and discuss the
impact of the different assumptions used in these analyses.

(3) We compare the TPE contributions as extracted using these different
approaches to more recent extractions~\cite{guttmann2011, borisyuk2011} which
attempt to extract the TPE amplitudes using additional constraints that come
from a recent measurement of the $\varepsilon$ dependence of ${G}_{Ep}/G_{Mp}$
at $Q^2=2.50$~(GeV/c)$^2$~\cite{meziane2011}.

\subsection{Form Factors and the TPE Amplitudes} \label{form_factors_tpe}

We begin with the extraction of the form factors and TPE contributions based
on the QA parametrization, Eq.~(\ref{eq:qattanfit2}). The results of this fit
for the data sets of Refs.~\cite{walker94, andivahis94, christy04, qattan05,
litt70, berger71, bartel73} are given in Table ~\ref{fit3}. The extraction
follows the procedure of fit III from Ref.~\cite{qattan2011}, with the fit to
the polarization transfer data, Eq.~(\ref{eq:ratio2}), used to constrain $R$,
leaving $G_{Mp}$ and $f$ as the two fit parameters. We also extract the TPE
contributions using the BK parametrization Eq.~(\ref{eq:Kobushkin3}).
Table~\ref{fitboriskoub} lists the results of the BK fit.

Note that for these older data sets, we take the cross sections used in
the analysis of Refs.~\cite{arrington03a, arrington04a}, where missing higher
older radiative corrections terms such as the Schwinger term and additional
vacuum polarization contributions from muon and quark loops have been applied.
See Ref.~\cite{note_resdata} for tabulated electron-proton elastic scattering
cross sections data.

\begin{table}[!htbp]
\begin{center}
\caption{The form factors and TPE parameters obtained using the QA
parametrization, Eq.~(\ref{eq:qattanfit2}), as a function of $Q^2$ (given
in units of (GeV/c)$^2$).  The TPE correction is extracted based on the
assumption that it fully resolves the difference between
$\mugegm$ as extracted from the given Rosenbluth extraction and the 
value from the polarization transfer parametrization of Eq.~(\ref{eq:ratio2}).
The overall normalization uncertainty for each data set (typically 1.5--3\%)
is not taken into account in this extraction.  See text for complete details.}
\begin{tabular}{c c c c c}
\hline \hline
$Q^2$ & $(G_{Mp}/\mu_{p}G_D)^{2}$& $(G_{Ep}/G_D)^{2}$& $f(Q^2)\times 100$ &$\chi^2_{\nu}$  \\
\hline
\hline
\multicolumn{5}{l}{Andivahis~\cite{andivahis94} (taken from Ref.~\cite{qattan2011})} \\
1.75 & 1.106$\pm$0.014& 0.669$\pm$0.009& 0.32$\pm$0.14 & 0.30 \\
2.50 & 1.111$\pm$0.013& 0.514$\pm$0.006& 0.07$\pm$0.04 & 0.53 \\
3.25 & 1.092$\pm$0.018& 0.371$\pm$0.006& 0.05$\pm$0.02 & 0.16 \\
4.00 & 1.068$\pm$0.017& 0.251$\pm$0.004& 0.02$\pm$0.01 & 0.51 \\
5.00 & 1.029$\pm$0.017& 0.130$\pm$0.002& 0.01$\pm$0.005& 0.93 \\
\hline
\multicolumn{5}{l}{Walker~\cite{walker94} (taken from Ref.~\cite{qattan2011})} \\
1.00 & 1.024$\pm$0.055& 0.785$\pm$0.042& 2.40$\pm$1.97 & 0.72 \\
2.00 & 1.020$\pm$0.037& 0.563$\pm$0.020& 0.67$\pm$0.18 & 0.65 \\
2.50 & 1.040$\pm$0.034& 0.483$\pm$0.016& 0.28$\pm$0.08 & 0.98 \\
3.00 & 1.010$\pm$0.045& 0.378$\pm$0.017& 0.18$\pm$0.06 & 0.19 \\
\hline
\multicolumn{5}{l}{Christy~\cite{christy04} (taken from Ref.~\cite{qattan2011})} \\
0.65 & 0.916$\pm$0.061& 0.777$\pm$0.052& 13.07$\pm$6.33& 0.02 \\
0.90 & 1.047$\pm$0.036& 0.826$\pm$0.028& 1.58$\pm$1.81 & 1.37\\
2.20 & 1.110$\pm$0.030& 0.574$\pm$0.016& 0.13$\pm$0.12 & 1.07\\
2.75 & 1.115$\pm$0.021& 0.468$\pm$0.009& 0.07$\pm$0.05 & 0.04\\
3.75 & 1.087$\pm$0.030& 0.291$\pm$0.008& 0.03$\pm$0.02 & 1.15\\
4.25 & 1.023$\pm$0.024& 0.210$\pm$0.005& 0.05$\pm$0.01 & 0.55\\
5.25 & 1.015$\pm$0.065& 0.106$\pm$0.007& 0.02$\pm$0.01 & 0.78\\
\hline
\multicolumn{5}{l}{Qattan~\cite{qattan05}} \\
2.64& 1.108$\pm$0.007& 0.486$\pm$0.003 & 6.39$\pm$1.10 & 0.35   \\
3.20& 1.098$\pm$0.008& 0.381$\pm$0.003 & 8.13$\pm$1.32 & 0.54   \\
4.10& 1.064$\pm$0.010& 0.237$\pm$0.002 & 10.80$\pm$1.78 & 0.14  \\
\hline
\multicolumn{5}{l}{Bartel~\cite{bartel73}} \\
1.169& 1.045$\pm$0.029& 0.761$\pm$0.021&   1.63$\pm$0.89 & 0.26  \\
1.75 & 1.103$\pm$0.029& 0.667$\pm$0.018&$-$0.03$\pm$0.30 & 0.01  \\
\hline
\multicolumn{5}{l}{Litt~\cite{litt70}} \\
1.50&  0.970$\pm$0.247& 0.637$\pm$0.162&   1.86$\pm$2.67 & 0.01  \\
2.00&  0.961$\pm$0.230& 0.534$\pm$0.128&   0.86$\pm$1.09 & 1.10  \\
2.50&  1.031$\pm$0.062& 0.477$\pm$0.029&   0.23$\pm$0.15 & 0.65  \\
3.75&  0.960$\pm$0.081& 0.257$\pm$0.022&   0.10$\pm$0.05 & 0.32  \\
\hline
\multicolumn{5}{l}{Berger~\cite{berger71}} \\
0.389& 0.970$\pm$0.037& 0.884$\pm$0.034&$-$0.9$\pm$12.8  & 0.48  \\
0.584& 0.970$\pm$0.017& 0.837$\pm$0.015&   5.27$\pm$2.78 & 0.38  \\
0.779& 1.008$\pm$0.025& 0.823$\pm$0.020&   1.82$\pm$2.22 & 0.55  \\
0.973& 1.005$\pm$0.032& 0.776$\pm$0.025&   3.37$\pm$1.65 & 0.81  \\
1.168& 1.047$\pm$0.048& 0.763$\pm$0.035&   2.35$\pm$1.94 & 0.29  \\
1.363& 1.076$\pm$0.046& 0.738$\pm$0.031&   0.29$\pm$1.22 & 0.58  \\
1.558& 1.063$\pm$0.057& 0.685$\pm$0.036&   1.71$\pm$1.48 & 1.12  \\
1.752& 1.129$\pm$0.077& 0.682$\pm$0.047&   0.07$\pm$1.21 & 0.75  \\
\hline \hline
\end{tabular}
\label{fit3}
\end{center}
\end{table}

\begin{table}[!htbp]
\begin{center}
\caption{The form factors and TPE parameters obtained using the BK
parametrization, Eq.~(\ref{eq:Kobushkin3}), as a function of $Q^2$ (given
in units of (GeV/c)$^2$).  The TPE correction is extracted based on the
assumption that it fully resolves the difference between
$\mugegm$ as extracted from the given Rosenbluth extraction and the 
value from the polarization transfer parametrization of Eq.~(\ref{eq:ratio2}).
The overall normalization uncertainty for each data set (typically 1.5--3\%)
is not taken into account in this extraction. See text for complete details.}
\bigskip
\begin{tabular}{c c c c c c}
\hline \hline
$Q^2$ & $(G_{Mp}/\mu_{p}G_D)^{2}$& $(G_{Ep}/G_D)^{2}$& $a(Q^2)\times 100$ & $\chi^2_{\nu}$&$N_{\mbox{p}}$ \\ \hline
\hline
\multicolumn{6}{l}{Andivahis~\cite{andivahis94}} \\
1.75&1.158$\pm$0.012& 0.700$\pm$0.007  & $-$2.25$\pm$0.94  & 0.30 & 4 \\
2.50&1.147$\pm$0.010& 0.531$\pm$0.004  & $-$1.57$\pm$0.88  & 0.53 & 7 \\
3.25&1.149$\pm$0.013& 0.390$\pm$0.004  & $-$2.48$\pm$1.16  & 0.16 & 5 \\
4.00&1.128$\pm$0.012& 0.265$\pm$0.003  & $-$2.64$\pm$1.10  & 0.51 & 6 \\
5.00&1.090$\pm$0.014& 0.138$\pm$0.002  & $-$2.82$\pm$1.24  & 0.93 & 5 \\
\hline
\multicolumn{6}{l}{Walker~\cite{walker94}} \\
1.00&1.101$\pm$0.012& 0.844$\pm$0.009  & $-$3.50$\pm$3.29  & 0.57 & 3 \\
2.00&1.183$\pm$0.009& 0.656$\pm$0.005  & $-$6.92$\pm$1.82  & 0.56 & 8 \\
2.50&1.177$\pm$0.010& 0.545$\pm$0.005  & $-$5.84$\pm$1.90  & 0.74 & 6 \\
3.00&1.178$\pm$0.014& 0.445$\pm$0.005  & $-$7.15$\pm$2.30  & 0.18 & 5 \\
\hline
\multicolumn{6}{l}{Christy~\cite{christy04}} \\
0.65&1.058$\pm$0.010& 0.897$\pm$0.009  & $-$6.71$\pm$3.18  & 0.01 & 3 \\
0.90&1.086$\pm$0.011& 0.856$\pm$0.009  & $-$1.77$\pm$2.10  & 1.37 & 3 \\
2.20&1.154$\pm$0.014& 0.597$\pm$0.007  & $-$1.92$\pm$1.84  & 1.07 & 3 \\
2.75&1.162$\pm$0.014& 0.487$\pm$0.006  & $-$2.04$\pm$1.34  & 0.04 & 3 \\
3.75&1.147$\pm$0.020& 0.307$\pm$0.005  & $-$2.61$\pm$2.03  & 1.47 & 3 \\
4.25&1.169$\pm$0.021& 0.240$\pm$0.004  & $-$6.22$\pm$1.64  & 0.55 & 3 \\
5.25&1.123$\pm$0.041& 0.117$\pm$0.004  & $-$4.83$\pm$4.50  & 0.78 & 3 \\
\hline
\multicolumn{6}{l}{Qattan~\cite{qattan05}} \\
2.64&1.174$\pm$0.006& 0.515$\pm$0.003  & $-$2.81$\pm$0.49  & 0.35 & 5 \\
3.20&1.183$\pm$0.007& 0.411$\pm$0.003  & $-$3.60$\pm$0.57  & 0.54 & 4 \\
4.10&1.176$\pm$0.011& 0.262$\pm$0.002  & $-$4.77$\pm$0.78  & 0.14 & 3 \\
\hline
\multicolumn{6}{l}{Bartel~\cite{bartel73}} \\
1.169&1.125$\pm$0.022& 0.819$\pm$0.016 & $-$3.57$\pm$1.89  & 0.26 &3 \\
1.75 &1.099$\pm$0.028& 0.665$\pm$0.017 & $+$0.18$\pm$2.21  & 0.01 &3 \\
\hline
\multicolumn{6}{l}{Litt~\cite{litt70}} \\
1.50 &1.157$\pm$0.022& 0.759$\pm$0.015 & $-$8.09$\pm$10.2  & 0.01 &3 \\
2.00 &1.169$\pm$0.023& 0.649$\pm$0.013 & $-$8.86$\pm$6.80  & 1.10 &4 \\
2.50 &1.143$\pm$0.011& 0.529$\pm$0.005 & $-$4.90$\pm$2.81  & 0.65 &9 \\
3.75 &1.154$\pm$0.016& 0.309$\pm$0.004 & $-$8.40$\pm$3.59  & 0.32 &3 \\
\hline
\multicolumn{6}{l}{Berger~\cite{berger71}} \\
0.389&0.966$\pm$0.011& 0.881$\pm$0.010 & $+$0.17$\pm$2.35  & 0.48 & 7 \\
0.584&1.014$\pm$0.009& 0.876$\pm$0.008 & $-$2.20$\pm$1.14  & 0.38 &14 \\
0.779&1.038$\pm$0.018& 0.848$\pm$0.014 & $-$1.48$\pm$1.80  & 0.55 & 6 \\
0.973&1.106$\pm$0.022& 0.854$\pm$0.017 & $-$4.55$\pm$2.16  & 0.81 & 5 \\
1.168&1.162$\pm$0.055& 0.846$\pm$0.040 & $-$4.95$\pm$4.04  & 0.29 & 4 \\
1.363&1.099$\pm$0.054& 0.753$\pm$0.037 & $-$1.01$\pm$4.20  & 0.58 & 4 \\
1.558&1.265$\pm$0.126& 0.815$\pm$0.081 & $-$8.01$\pm$6.26  & 1.12 & 3 \\
1.752&1.141$\pm$0.135& 0.690$\pm$0.082 & $-$0.54$\pm$9.20  & 0.75 & 3 \\
\hline \hline
\end{tabular}
\label{fitboriskoub}
\end{center}
\end{table}

\begin{figure}[!htbp]
\begin{center}
\includegraphics*[width=8.3cm]{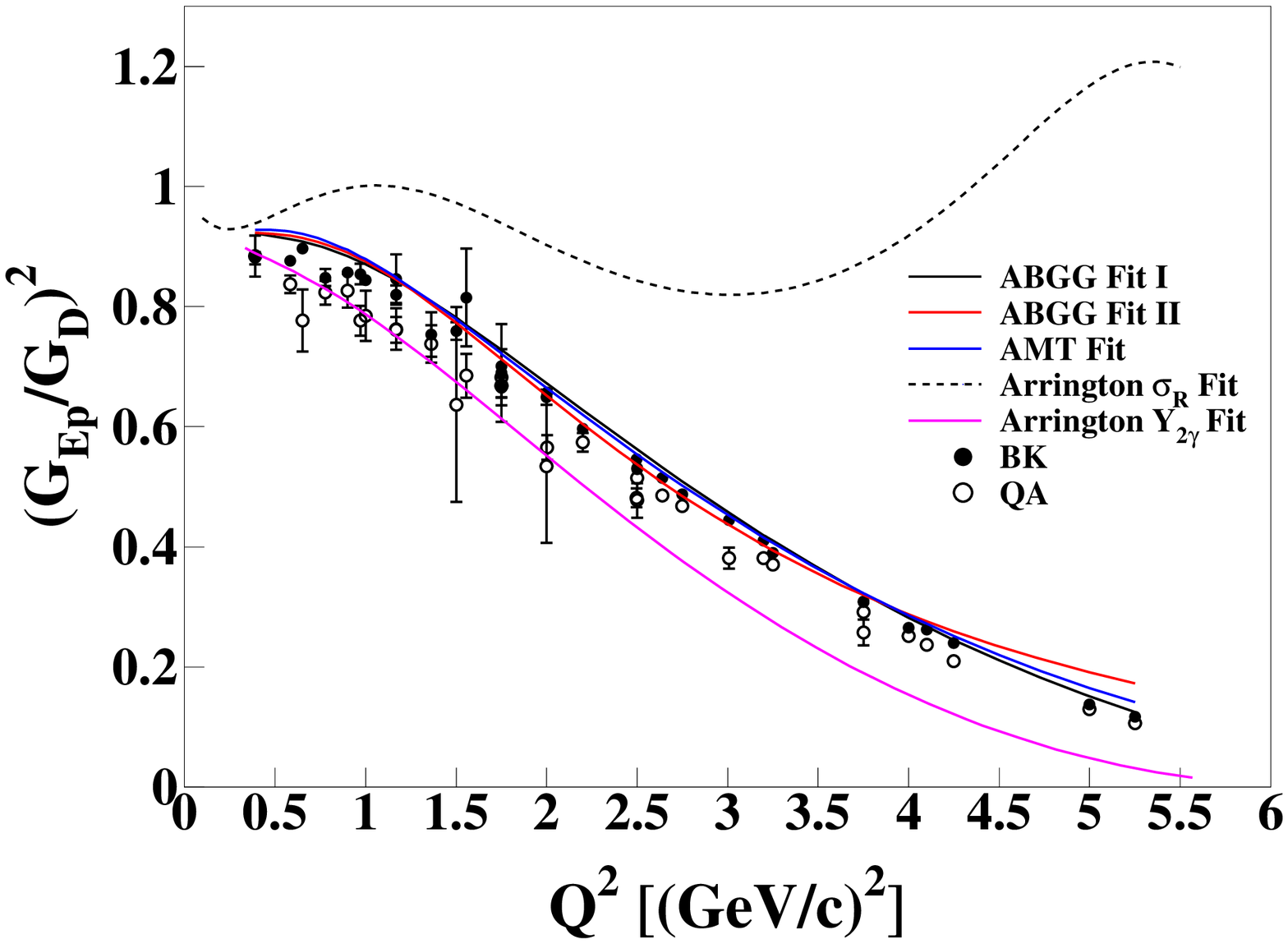}\\
\includegraphics*[width=8.3cm]{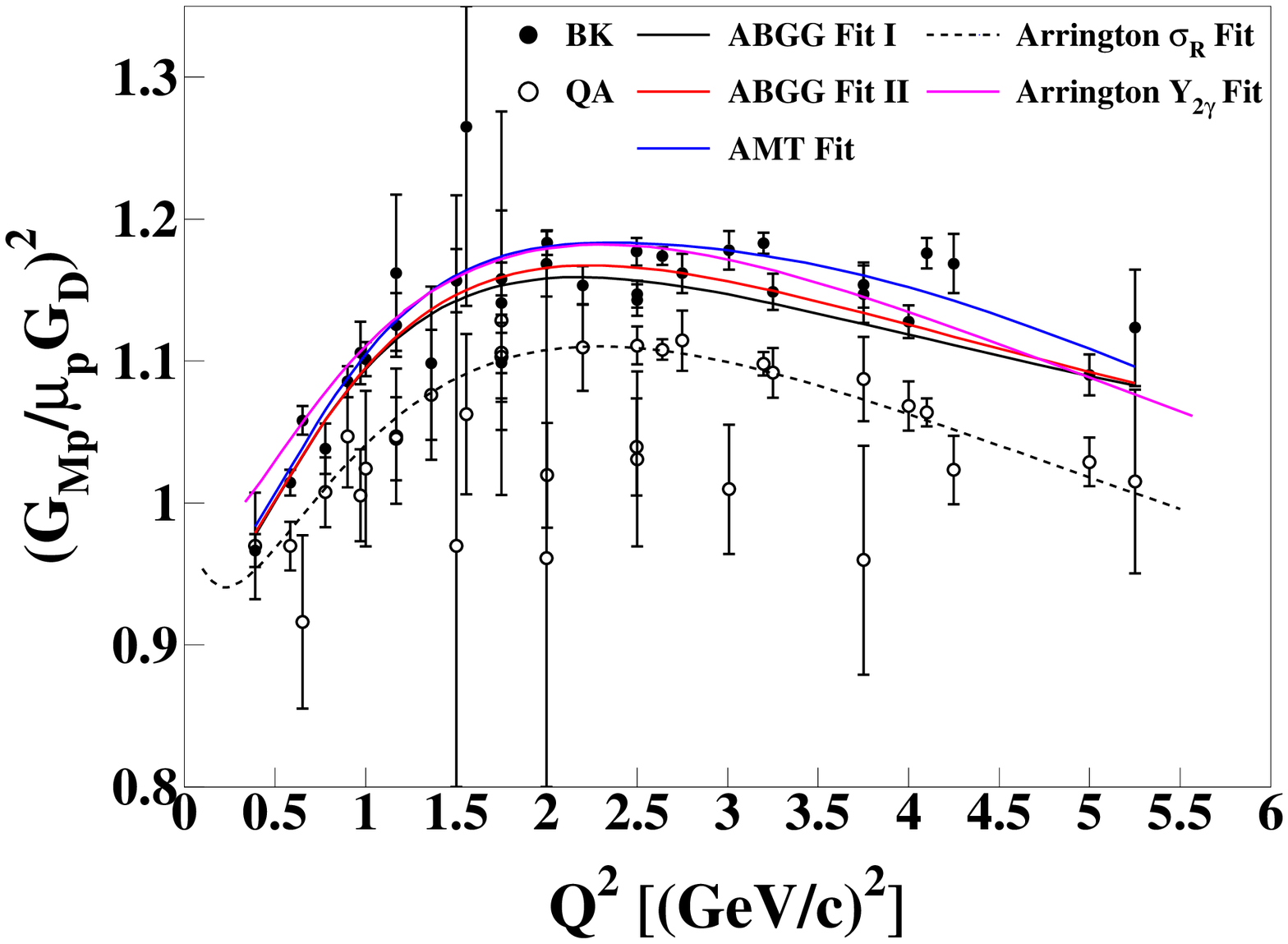}
\end{center}
\vspace{-0.5cm}
\caption{(Color online) $(G_{Ep}/G_{D})^2$ [top] and $(G_{Mp}/\mu_{p}G_{D})^2$
[bottom] as obtained from Refs.~\cite{andivahis94, qattan05, walker94,
christy04, bartel73, litt70, berger71} using the QA and BK parametrizations.
Also shown are the ABGG I and ABGG II fits~\cite{alberico09a}, the AMT
fit~\cite{arrington07}, the Arrington $Y_{2\gamma}$ Fit~\cite{arrington05},
and the Arrington $\sigma_{R}$ Fit~\cite{arrington04a} which does not include
any TPE contributions or polarization data.}
\label{fig:Ge2}
\end{figure}

Figure~\ref{fig:Ge2} shows the values of $(G_{Ep}/G_{D})^{2}$ and $(G_{Mp}/
\mu_{p} G_{D})^{2}$ obtained using the QA and BK parametrizations (shown as
hollow and solid circles, respectively).  The solid curves are fits to the
extracted form factors from previous phenomenological analyses: 
Ref.~\cite{alberico09a} (ABGG Fit I and ABGG Fit II), and
Ref.~\cite{arrington05} (Arrington $Y_{2\gamma}$ Fit), and the form factors
extracted based on a calculation of the TPE contributions from
Ref.~\cite{arrington07} (AMT Fit).  The dotted line is the global Rosenbluth
extraction from Ref.~\cite{arrington04a} (Arrington $\sigma_R$ Fit), which
does not include TPE corrections.

The QA results for $G_{Mp}$ are typically 2--10\% below the BK values.
This simply reflects the fact that the QA parametrization yields no TPE
contribution at $\varepsilon=0$, and thus the (linear) extrapolation to
the \textit{anchor point} at $\varepsilon=0$ is unchanged by the application
of the extracted TPE contributions, while the extrapolation is modified by the
TPE contributions for the BK parametrization.  Note that because they use the
same PT parametrization for $\mugegm$, the $\gep$ values are also lower by
the same amount for the QA extraction.  In addition, the uncertainties are
much larger in the QA extraction for some kinematics.  Most experiments tend
to have a large part of their data at large $\varepsilon$ values, so there is a
significant extrapolation to $\varepsilon=0$ in the QA extraction, yielding a
larger uncertainty in the overall normalization.  For the BK parametrization,
the normalization is fixed at the $\varepsilon=1$ value. This reduction in the
uncertainties demonstrates one of the important strengths of the polarization
transfer measurements, as they significantly reduce the uncertainty associated
with the extraction to $\varepsilon=0$.  While the uncertainty on the PT data
is small, neglecting this uncertainty in the fit (as we and some other
extractions do), yields a small underestimate of the uncertainty.  The
uncertainties associated with this, and other assumptions that go into the
phenomenological extractions, will be discussed in Sec.~\ref{Rpm_Ratio}.

In the extraction of $(G_{Ep}/G_{D})^{2}$, there is a large difference between
the TPE-uncorrected result (Arrington $\sigma_R$) and the extractions that
apply corrections for TPE contributions.  This is because the LT extraction of
$\gep$ at large $Q^2$ is extremely sensitive to angular-dependent TPE
corrections, since $\gep$ enters into the cross section as a small,
$\varepsilon$-dependent term. Nearly all of the other fits extractions are in
excellent agreement, but this is simply because they all explain the
difference between the LT and PT extractions of $\gegm$ in terms of a
nearly-linear correction to the reduced cross section.  Therefore, these
analyses will, by construction, ensure that the final result for $\gegm$ will
be consistent with the polarization data, thus yielding nearly identical
values for $\gep$ except for small differences in the extracted values of
$(G_{Mp}/\mu_pG_{D})^{2}$, most clearly seen in the QA extraction. The quality
of the agreement shows that the small non-linearities in the ABGG analysis and
the choice of cross section data sets and parametrization for the polarization
transfer measurements has little impact.  The AMT analysis applies TPE
contributions based on a hadronic calculation~\cite{blunden05a}, with a small
additional contribution designed to more fully resolve the discrepancy for
$Q^2 > 2$~(GeV/c)$^2$, where the calculated TPE do not bring the LT and PT
results into perfect agreement. Because of this additional contribution, and
the neglect of TPE contributions to the polarization data, this analysis must
also yield results consistent with the other analyses at large $Q^2$.  The
only analysis that yields different results is the Arrington $Y_{2\gamma}$ fit,
where the extraction of all three amplitudes allows for a contribution of the
TPE contribution to the polarization transfer data, yielding a noticeable
downward shift in the value of $\mugegm$ and thus $(G_{Ep}/G_{D})^{2}$. 
However, the size and even the sign of this correction depend on the
assumptions made in trying to separate the three TPE amplitudes.  Thus, the
deviation from the other fits is at best an indication of the possible
uncertainty in these extractions. This will be addressed further when we
compare to other extractions that include TPE contributions to the
polarization data in Sec.~\ref{Rpm_Ratio}.

For the extraction of $(G_{Mp}/ \mu_{p} G_{D})^{2}$, the BK results are in
good agreement with previous extractions that include TPE, while the QA
results are more consistent with the Arrington $\sigma_R$ fit which neglects
TPE.  The BK and ABGG other extractions assume no TPE contribution to the
polarization data and an approximately linear correction to the reduced cross
section, as discussed above.  Because they require very similar corrections to
the $\varepsilon$ dependence of $\sigma_R$ to resolve the discrepancy in
$\mugegm$ measurements, they have very similar corrections to $(G_{Mp}/
\mu_{p} G_{D})^{2}$, which simply relates to the low-$\varepsilon$ value of
the correction. The Arrington $Y_{2\gamma}$ fit is the only extraction in
which the polarization transfer results for $G_{Ep}/G_{Mp}$ are modified by
TPE contributions.  While the overall size of the TPE correction required to
resolve the LT-PT discrepancy is somewhat larger in this case, the absolute
correction to the cross section is small, yielding very similar values for
$(G_{Mp}/ \mu_{p} G_{D})^{2}$.

\begin{figure}[!htbp]
\begin{center}
\includegraphics*[width=8.6cm]{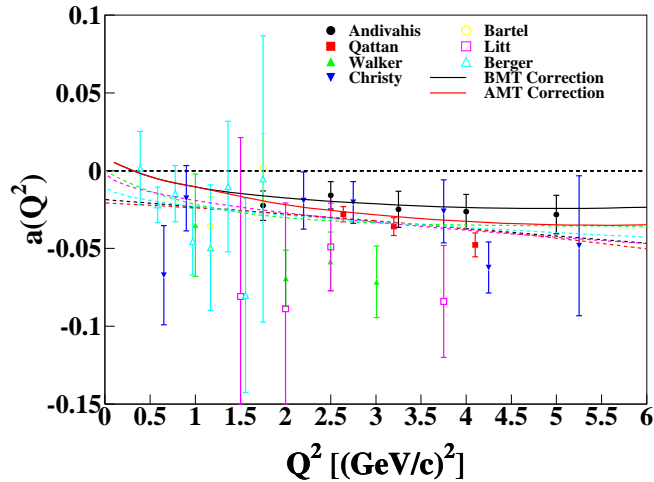}
\end{center}
\vspace{-0.5cm}
\caption{(Color online) The fit parameter $a(Q^{2})$ as obtained using the BK
parametrization from the data of
Refs.~\cite{andivahis94,qattan05,walker94,christy04,bartel73,litt70,berger71}.
The dashed curves correspond to fits to the $Q^2$ dependence using different
parametrizations (see text) and the solid lines correspond to the values of
$a(Q^2)$ determined by fitting to the TPE calculations of
Ref.~\cite{blunden05a}, and the TPE correction of Ref.~\cite{arrington07}
which adds and additional phenomenological TPE contribution at higher $Q^2$.}
\label{fig:aQ2}
\end{figure}

Figure~\ref{fig:aQ2} shows the fit parameter $a(Q^2)$ as a function of $Q^{2}$
for all data sets. The parameter $a(Q^2)$ is at the few percent level, and for
the most part, increases in magnitude with increasing $Q^{2}$.  Note that the
TPE contribution to the cross section is $F(Q^2,\varepsilon) = 2 a
(1-\varepsilon) G_{Mp}^2$, and thus at high $Q^2$, the fractional slope
introduced by the TPE correction is $2a$.  The extracted TPE contributions
from the different data sets are relatively consistent, and show a slow
increase in the TPE contribution as $Q^2$ increases.

In an attempt to parametrize the $Q^2$ dependence of the parameter $a(Q^2)$,
several different functional forms were tried and are plotted as dashed lines
in Fig.~\ref{fig:aQ2}.  These fits all give reasonable reduced $\chi^2$
values, 1.04 $<\chi_{\nu}^2<$ 1.10.  The lowest $\chi_{\nu}^2$ value was
obtained using the form $a(Q^2) = \alpha\sqrt{Q^2}$ with
$\alpha$=$-$0.0191$\pm$0.0014 (magenta dashed line in Fig.~\ref{fig:aQ2}). 
The full spread of the fits is below 0.005 for 1$<$$Q^2$$<$4~(GeV/c)$^2$,
although at lower $Q^2$ values, none of the extractions are precise and the
behavior is largely unconstrained below $Q^2 \approx 1.5$~(GeV/c)$^2$. 
For our global fit, we take
\begin{equation}\label{eq:globfit}
a(Q^2) = -0.0191\sqrt{Q^2} \pm 0.0014\sqrt{Q^2} \pm 0.003
\end{equation}
where the first uncertainty is the fit uncertainty in $\alpha$, and the second
is the systematic error band included to account for the model dependence of
the fit for $1.5<Q^2<4$~(GeV/c)$^2$.

Figure~\ref{fig:aQ2} also shows curves for $a(Q^2)$ based on the hadronic TPE
correction of Ref.~\cite{blunden05a}, along with the version used in
Ref.~\cite{arrington07} which includes a small additional contribution at high
$Q^2$ values.  The version used in the global analysis of the form
factors~\cite{arrington07} is in good agreement with our fit for $Q^2 \gtorder
2$~(GeV/c)$^2$, where the data provides significant constraints on the TPE
contributions.  Note that the calculated TPE corrections, as well as those of
Refs.~\cite{arrington04c, borisyuk07}, show a change of sign in the TPE
effects for $Q^2<0.5$~(GeV/c)$^2$.  This is not seen in the data, although the
constraints at low $Q^2$ are insufficient to make strong conclusions in this
region.

The fact that the low $Q^2$ behavior is inconsistent with the calculations
is not entirely surprising.  Our extraction, like most similar
phenomenological analyses, assumes that TPE contributions are significant
for the cross section measurements but negligible for polarization data. 
However, both the hadronic~\cite{blunden05a} and partonic~\cite{afanasev05a}
calculations suggest that the TPE contributions are at the few percent level
for both observables. The main difference is that the impact of the correction
on the extracted form factors is amplified for $\gep$ at high $Q^2$, where the
form factor is extracted from a small angular dependence in the cross section
which can be noticeably modified by a few percent TPE contribution.  At low
$Q^2$, the TPE contribution is not amplified in the extraction of the form
factors from the cross section data, so neglecting the contributions to the
polarization transfer data will not be reliable.  In addition, because the
difference between $\mugegm$ from Rosenbluth and polarization transfer
measurements becomes small at low $Q^2$, neglecting the uncertainty in the
polarization transfer extraction of $\mugegm$ leads to a significant
underestimate of the uncertainties below 1~(GeV/c)$^2$.

\begin{figure}[!htbp]
\begin{center}
\includegraphics*[width=8.6cm]{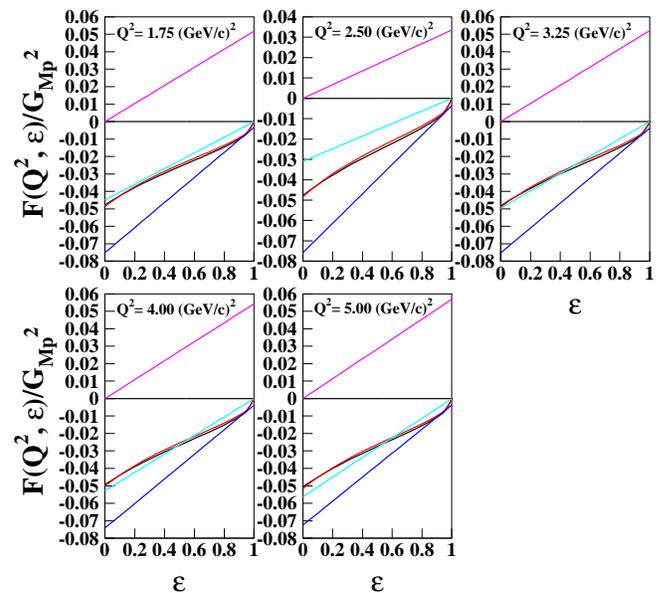}
\end{center}
\vspace{-0.5cm}
\caption{(Color online) The Ratio $F(Q^2,\varepsilon)/G_{Mp}^2$ as a function of $\varepsilon$ at 
$Q^2=$ 1.75, 2.50, 3.25, 4.0, and 5.0 (GeV/c)$^2$ from Ref.~\cite{andivahis94} determined using the QA
parametrization (magenta line), the BK parametrization (cyan line), the ABGG Fit I (black line) and the 
ABGG Fit II (red line), and the Arrington $Y_{2\gamma}$ Fit (blue line).}
\label{fig:TPEcorr}
\end{figure}

In Fig.~\ref{fig:TPEcorr}, we compare the $\varepsilon$ dependence of the
TPE corrections based on the different extractions based on the data
from Ref.~\cite{andivahis94}.  In addition to the extractions from the QA and
BK parametrizations, we show the ABGG Fit I (black line), ABGG Fit II (red
line), the Arrington $Y_{2\gamma}$ Fit (blue line). Most of the extractions
yield similar slopes, with the QA extraction differing from the others in that
the TPE contribution goes to zero at $\varepsilon=0$.  The Arrington
$Y_{2\gamma}$ parametrization has a somewhat larger slope, because of the
inclusion of TPE contributions which reduce $R$ as measured in polarization
experiments, thus necessitating a greater decrease to $R$ as extracted in a
Rosenbluth separation.  The TPE corrections increase slowly with $Q^2$ except
for the ABGG fits which show essentially no $Q^2$ dependence.  In these
extractions, the $Q^2$ dependence of the TPE contribution is taken to 
go as $G_D^2(Q^2)$, and thus they are a nearly constant fractional correction
to the cross section at large $Q^2$ values, where $G_M \approx \mu_p G_D$
dominates the cross section.

\subsection{The $R_{e^{+} e^{-}}$ Ratio} \label{Rpm_Ratio}

The function $F(Q^2,\varepsilon)$ which represents the interference of the OPE
and TPE amplitudes, changes sign depending on the charge of the projectile,
yielding an amplified signal when taking the ratio of electron and positron
scattering. The ratio $R_{e^{+} e^{-}}(Q^2,\varepsilon)$, defined by Eq.
(\ref{eq:ratiopostelect}), is determined simply by changing the sign in front
of the TPE amplitudes. We determined the ratio $R_{e^{+} e^{-}}$ using the
form factors and the TPE amplitude extracted using the QA and BK
parametrizations. These ratios will then be compared to those obtained from
other analyses~\cite{alberico09a, arrington05, guttmann2011, borisyuk2011},
and calculations~\cite{arrington04c, blunden05a, arrington07}.  We also 
use our extracted TPE contribution to make predictions for new and recently
completed measurements~\cite{vepp_proposal, e07-005, nikolenko2010a,
nikolenko2010b, kohl09}.

\begin{figure}[!htbp]
\begin{center}
\includegraphics*[width=8.6cm]{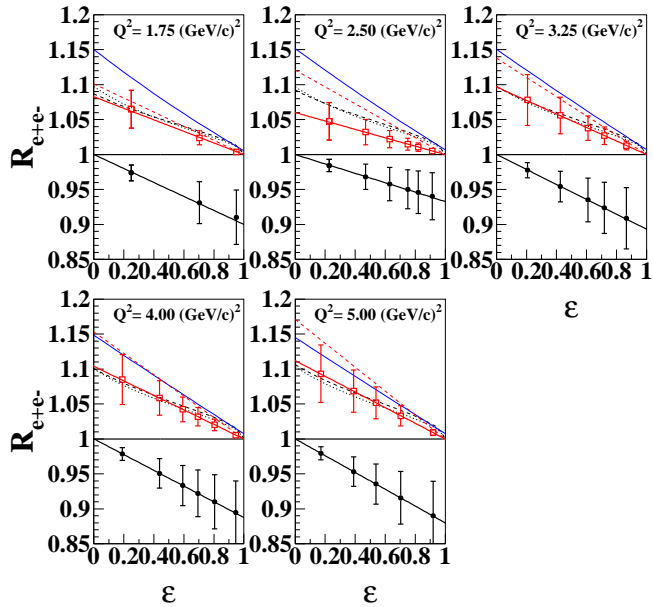}
\end{center}
\vspace{-0.5cm}
\caption{(Color online) $R_{e^{+} e^{-}}$ as a function of $\varepsilon$ at
$Q^2=$ 1.75, 2.50, 3.25, 4.0, and 5.0 (GeV/c)$^2$ from Ref.~\cite{andivahis94}
determined using the QA parametrization (solid black circles), and the BK
parametrization (open red squares). The solid black and red lines through the
data are calculated using Eqs. (\ref{eq:fittedRpmQA}) and
(\ref{eq:fittedRpmBK}), respectively, while the dashed red line is the global
fit from the BK extraction Eq.~(\ref{eq:globfit}). Also shown are results of
the Arrington $Y_{2\gamma}$ Fit (solid blue line), the ABGG Fit I (black
dashed line) and the ABGG Fit II (black dotted line).}
\label{fig:RpmAndiFitII}
\end{figure}

Figure~\ref{fig:RpmAndiFitII} shows the ratio $R_{e^{+} e^{-}}$ as a function
of $\varepsilon$ extracted from the data of Ref.~\cite{andivahis94} using
the QA parametrization (solid black circle) and the BK parametrization (open
red squares), along with the previous extractions.  The ratio
$R_{e^{+} e^{-}}$ as determined using the QA parametrization is always equal
to or less than unity, while the other parametrizations force $R_{e^{+}
e^{-}}=0$ at $\varepsilon=1$, yielding a ratio that is always equal to or
larger than one. Because measurements of $R_{e^{+} e^{-}}$ put significant
constraints on deviations from unity at $\varepsilon=1$, we exclude the QA
parametrization from further comparisons. As seen in Figs.~\ref{fig:TPEcorr}
and~\ref{fig:RpmAndiFitII}, the $\varepsilon$ dependence is very similar
to that of the BK fit, except for the overall offset.

\begin{figure}[!htbp]
\begin{center}
\includegraphics*[width=8.3cm]{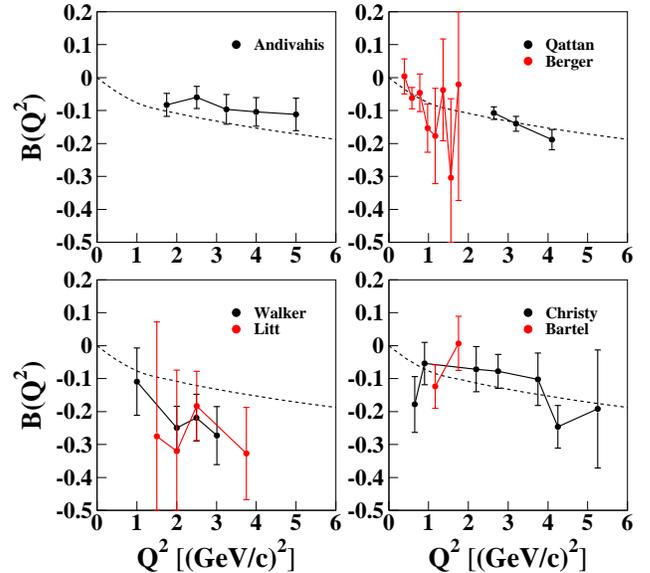}
\end{center}
\vspace{-0.5cm}
\caption{(Color online) The slope $B(Q^2)$ as a function of $Q^2$ as
determined using the BK parametrization. The solid lines connecting the data
points are to guide the eye, while the black dashed curve is our global best
fit.}
\label{fig:RpmSlope}
\end{figure}

For a better comparison between the two parametrizations, we fit the ratio
extracted using the QA parametrization to the form
\begin{equation}
\label{eq:fittedRpmQA} R_{e^{+} e^{-}}(Q^2,\varepsilon) = 1+B(Q^2)\varepsilon,
\end{equation}
and that extracted using the BK parametrization to
\begin{equation}
\label{eq:fittedRpmBK} R_{e^{+} e^{-}}(Q^2,\varepsilon) = 1-B(Q^2)(1-\varepsilon),
\end{equation}
with $B(Q^2)$ being the parameter of the fit and represents the slope.
The values of $B(Q^2)$ from the BK fit are given in
Table~\ref{fitRpmfitboriskoub} and shown as a function of $Q^2$ for each
experiment in Fig.~\ref{fig:RpmSlope}. The slope $B(Q^2)$ is negative and
grows in magnitude with increasing $Q^2$ value almost for most data sets. We
also compare the results to a reference curve corresponding to our global fit
Eq.~(\ref{eq:globfit}).

\begin{table}[!htbp]
\begin{center}
\caption{The values of the slope $B(Q^{2})$ as obtained using the QA and BK
parametrizations by fitting to Eqs. (\ref{eq:fittedRpmQA}) and
(\ref{eq:fittedRpmBK}), respectively.}
\begin{tabular}{c c c} 
\hline \hline
$Q^2$(GeV/c)$^2$  &  $B(Q^{2})$ [QA]  &  $B(Q^2)$ [BK] \\
\hline
&& Andivahis~\etal (Ref.~\cite{andivahis94}) \\ 
1.75&$-$0.0999$\pm$0.0434	&	$-$0.0827$\pm$0.0346   \\
2.50&$-$0.0672$\pm$0.0378	&	$-$0.0600$\pm$0.0337   \\
3.25&$-$0.1067$\pm$0.0513	&	$-$0.0967$\pm$0.0451   \\
4.00&$-$0.1124$\pm$0.0483	&	$-$0.1037$\pm$0.0434   \\
5.00&$-$0.1200$\pm$0.0537	&	$-$0.1119$\pm$0.0494   \\ \hline
&& Walker~\etal (Ref.~\cite{walker94})  \\
1.00&$-$0.1571$\pm$0.1291	&	$-$0.1090$\pm$0.1022   \\
2.00&$-$0.3268$\pm$0.0874	&	$-$0.2499$\pm$0.0656   \\
2.50&$-$0.2681$\pm$0.0811	&	$-$0.2186$\pm$0.0710   \\
3.00&$-$0.3378$\pm$0.1142	&	$-$0.2729$\pm$0.0878   \\ \hline
&& Christy~\etal (Ref.~\cite{christy04})   \\
0.65&$-$0.3274$\pm$0.1588	&	$-$0.1784$\pm$0.0846   \\
0.90&$-$0.0786$\pm$0.0904	&	$-$0.0542$\pm$0.0645   \\
2.20&$-$0.0813$\pm$0.0744	&	$-$0.0713$\pm$0.0684   \\
2.75&$-$0.0871$\pm$0.0581	&	$-$0.0781$\pm$0.0513   \\
3.75&$-$0.1116$\pm$0.0880	&	$-$0.1022$\pm$0.0796   \\
4.25&$-$0.2872$\pm$0.0791	&	$-$0.2461$\pm$0.0648   \\
5.25&$-$0.2141$\pm$0.2037	&	$-$0.1921$\pm$0.1790   \\ \hline
&& Qattan~\etal (Ref.~\cite{qattan05})  \\
2.64&$-$0.1228$\pm$0.0212	&	$-$0.1078$\pm$0.0187   \\
3.20&$-$0.1585$\pm$0.0258	&	$-$0.1401$\pm$0.0223   \\
4.10&$-$0.2135$\pm$0.0352	&	$-$0.1886$\pm$0.0307   \\ \hline
&& Bartel~\etal (Ref.~\cite{bartel73})  \\
1.169&$-$0.1707$\pm$0.0933	&	$-$0.1239$\pm$0.0658 \\
1.750&$+$0.0079$\pm$0.0935	&	$+$0.0068$\pm$0.0820 \\ \hline
&& Litt~\etal (Ref.~\cite{litt70})    \\
1.50 &$-$0.3909$\pm$0.5630	&	$-$0.2748$\pm$0.3471   \\
2.00 &$-$0.4382$\pm$0.5581	&	$-$0.3202$\pm$0.2457   \\
2.50 &$-$0.2200$\pm$0.1404	&	$-$0.1830$\pm$0.1050   \\
3.75 &$-$0.4059$\pm$0.1935	&	$-$0.3271$\pm$0.1400   \\ \hline      
&& Berger~\etal (Ref.~\cite{berger71}) \\
0.389&$+$0.0079$\pm$0.1108	&	$+$0.0039$\pm$0.0534     \\
0.584&$-$0.1087$\pm$0.0579	&	$-$0.0623$\pm$0.0324     \\
0.779&$-$0.0705$\pm$0.0864	&	$-$0.0464$\pm$0.0572     \\
0.973&$-$0.2291$\pm$0.1121	&	$-$0.1534$\pm$0.0731     \\                           
1.168&$-$0.2516$\pm$0.2075	&	$-$0.1769$\pm$0.1447     \\
1.363&$-$0.0462$\pm$0.1942	&	$-$0.0374$\pm$0.1545     \\
1.558&$-$0.4284$\pm$0.3530	&	$-$0.3037$\pm$0.2395     \\
1.752&$-$0.0232$\pm$0.3807	&	$-$0.0205$\pm$0.3527     \\
\hline \hline
\end{tabular}
\label{fitRpmfitboriskoub}
\end{center}
\end{table}

\begin{figure}[!htbp]
\begin{center}
\includegraphics*[width=8.1cm]{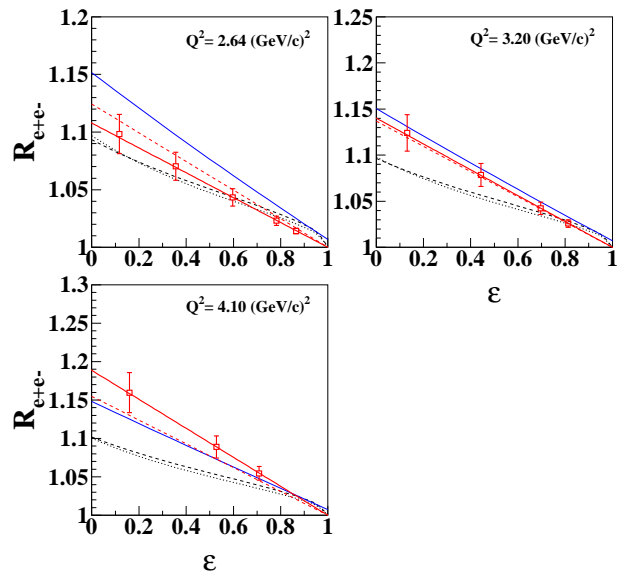}
\end{center}
\vspace{-0.5cm}
\caption{(Color online) $R_{e^{+} e^{-}}$ as a function of $\varepsilon$ as
extracted from Ref.~\cite{qattan05} using the BK parametrization (square
points and solid red line).  The red dashed lines show the global fit of
Eq.~(\ref{eq:globfit}).  Also shown are the results of the Arrington
$Y_{2\gamma}$ Fit (solid blue line), the ABGG Fit I (black dashed line) and
the ABGG Fit II (black dotted line).}
\label{fig:RpmQattanFitII}
\end{figure}

In Figs.~\ref{fig:RpmAndiFitII} and~\ref{fig:RpmQattanFitII}, we compare
previous TPE extractions to our results for the kinematics of the two highest
precision data sets~\cite{andivahis94, qattan05}. The ratio $R_{e^{+} e^{-}}$
as predicted by ABGG Fit 1 and ABGG Fit 2 are nearly identical, but smaller
than the Arrington $Y_{2\gamma}$ Fit.  For the Andivahis data, our extraction
is in generally good agreement with the ABGG fits, but the Qattan data yield
larger TPE corrections, especially at their higher $Q^2$ values.

At present, precise measurements of $R_{e^{+} e^{-}}$ are limited to
relatively low $Q^2$ or large $\varepsilon$, making it difficult to directly
compare the data to estimates of the $\varepsilon$ dependence of TPE at high
$Q^2$ extracted from comparisons of the Rosenbluth and polarization
measurements. As stated in Ref.~\cite{arrington04b}, if TPE corrections are
responsible for the discrepancy between the Rosenbluth and recoil-polarization
data, a 5--8\%, linear or quasi-linear, $\varepsilon$-dependent correction to
the electron cross section is required to resolve the discrepancy at high
$Q^2$.  This implies that $R_{e^{+} e^{-}}$ at high $Q^2$ should have a
10--16\% quasi-linear $\varepsilon$ dependence, with $R_{e^{+} e^{-}}$
decreasing as $\varepsilon$ increases.  The existing $R_{e^{+} e^{-}}$ data
provide some evidence for such a $\varepsilon$ dependence, but the
significance is only 3$\sigma$ and limited to low $Q^2$ data, where the
$\varepsilon$-dependence is observed to be
(5.7$\pm$1.8)\%~\cite{arrington04b}.  This corresponds to $B(Q^2)=0.057$, but
the extraction uses only data below $Q^2=2$~(GeV/c)$^2$, corresponding to an
average $Q^2$ values of approximately 0.5~(GeV/c)$^2$, making the result
consistent with the low $Q^2$ extractions presented here.  Relatively 
precise data do exist at large $\varepsilon$ and moderate $Q^2$ values,
suggesting that $R_{e^{+} e^{-}} \approx 1$ for $\varepsilon \to 1$, as
assumed in most of the extractions.

\begin{figure}[!htbp]
\begin{center}
\includegraphics*[width=8.1cm]{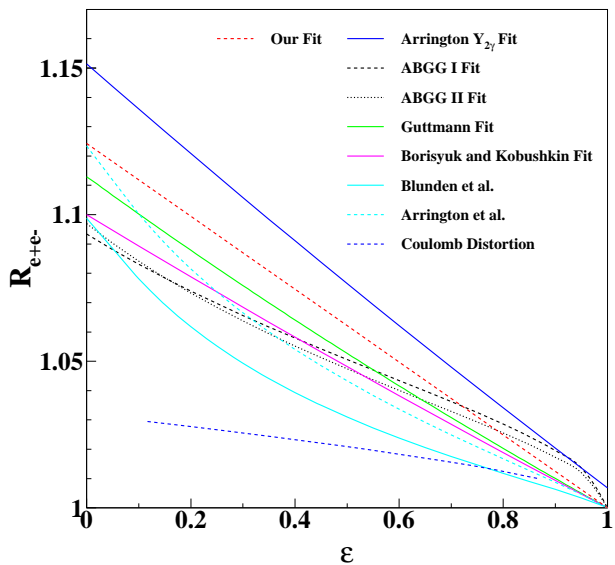}
\end{center}
\vspace{-0.5cm}
\caption{(Color online) Ratio $R_{e^{+} e^{-}}$ as a function of $\varepsilon$
at $Q^2=2.64$~(GeV/c)$^2$. In addition to the curves shown in
Fig.~\ref{fig:RpmQattanFitII}, we also show the extractions from
Guttmann~\etal~\cite{guttmann2011} and Borisyuk and
Kobushkin~\cite{borisyuk2011}, and calculations of Coulomb
distortion~\cite{arrington04c} and hadronic TPE contributions from Blunden
\etal~\cite{blunden05a}, and with the addition of a small additional
contribution at higher $Q^2$ values~\cite{arrington07}.  Note that the
extraction of Ref.~\cite{borisyuk2011} was performed at
$Q^2=2.50$~(GeV/c)$^2$.  The uncertainties in our fit Eq.~(\ref{eq:globfit})
correspond to an uncertainty of $\pm$0.015 at $\varepsilon$=0.}
\label{fig:TPEcompare}
\end{figure}

Finally, we make an additional comparison of the $\varepsilon$ dependence at
$Q^2=2.64$~(GeV/c)$^2$, including two recent extractions which take advantage
of new measurements of the $\varepsilon$ dependence of the recoil polarization
data~\cite{meziane2011}.  These are similar to the earlier attempts to 
separate the individual TPE amplitudes~\cite{guichon03, arrington05}, but
because of the new data on the $\varepsilon$ dependence of the polarization
measurement of $\mugegm$ and the polarization component $P_l$, fewer 
assumptions have to be made as to the contribution of the different amplitudes.

Guttmann~\etal~\cite{guttmann2011} used the measurements of the
$\varepsilon$ dependence of $P_l$ and $P_t/P_l$~\cite{meziane2011} along
with the $\varepsilon$ dependence of the cross section~\cite{qattan05} to
constrain the three TPE amplitudes $Y_M$, $Y_E$, and $Y_3$. By writing the
observables in terms of Born values and the TPE amplitudes and keeping only
the dominant TPE corrections (i.e. neglecting those that are suppressed by
kinematic or other factors), they observe that the corrections to
$P_l/P_l^{Born}$ are largely determined by $Y_3$ and the corrections to
$P_t/P_l$ (which shows no $\varepsilon$-dependence) come from a sum of $Y_3$
and $Y_E$.  Thus, the $\varepsilon$ dependence of $P_l$ is used to determine
$Y_3$, taking two different $\varepsilon$ dependences based on either
constraints from a pQCD calculation~\cite{borisyuk09} or a modified
parameterization. The lack of TPE contribution to $P_t/P_l$ is then
used to constrain constrains $Y_E$.  This allows $Y_M$ to be extracted based
on the difference between the predicted and measured values of the reduced
cross section as a function of $\varepsilon$.  This accounts for all of the
uncertainty in the extraction of the PT result for $\mugegm$ based on the
assumption that there is no $\varepsilon$-dependent TPE correction, but does
not account for the fact that the data are still consistent with a
$\varepsilon$ dependence of a few percent over the full $\varepsilon$ range.
Their extracted amplitudes are used to determine $R_{e^{+} e^{-}}$ at
$Q^2=2.64$~(GeV/c)$^2$, which reaches a maximum value of $1.11\pm0.016$,
is compared to the other extractions in Fig.~\ref{fig:TPEcompare}. 

The analysis of Borisyuk and Kobushkin~\cite{borisyuk2011} takes a similar
approach, although they use a different linear combination of amplitudes than
Ref.~\cite{guichon03}.  Again, the $\varepsilon$ dependence of $P_t/P_l$ is
taken to be zero~\cite{meziane2011}, and the correction to the cross section
is taken to be linear~\cite{tvaskis06}. Data from available electron-proton
scattering cross sections in the range of 2.20 $\le Q^2 \le$ 2.80 (GeV/c)$^2$
were interpolated to $Q^2=2.5$~(GeV/c)$^2$ to extract the amplitudes at a fixed
$Q^2$ value.  This yields an extraction in terms of a single amplitude, and
using the parametrization of Eq.~(\ref{eq:Kobushkin3}), they obtain a value
of $a=-0.0250\pm0.0035$, corresponding to a peak $R_{e^{+} e^{-}}$ value of
$1.11\pm0.016$.

Note that the values in our result and both of these extractions of the TPE
amplitudes are very similar, as is our uncertainty compared to the
Guttmann~\etal result.  While the details of the assumptions made in
separating the TPE amplitudes affect the prediction they make for polarization
observables, they both assume that there is no TPE contribution to the
polarization transfer ratio, and thus are extracting amplitudes designed to
induce a linear correction to the reduced cross section to make it consistent
with the PT data.  So for the observables related to the form factor
extraction, this is the same assumption made in all of the phenomenological
extractions except for the Arrington $Y_{2\gamma}$ fit.  Similarly, these
extractions typically neglect some of the uncertainties associated with the
TPE contributions to the PT measurements.  All of them ignore the uncertainty
in the extracted value of $\mugegm$ or the impact of a possible $\varepsilon$
dependence, and most of the phenomenological extractions ignore both.  The
Arrington $Y_{2\gamma}$ fit includes the uncertainty in the extraction of
$\mugegm$ from PT data, but not the uncertainty associated with the TPE.  One
can take the difference between the Arrington $Y_{2\gamma}$ fit and the other
extractions of $\gep$, Fig.~\ref{fig:Ge2}, as an estimate of the latter
uncertainty, but the TPE impact on the PT data in this analysis is rather
large, so this difference is probably better treated as an upper limit in the
uncertainty, at least for $Q^2$ up to 3--4~(GeV/c)$^2$.  A shift in the PT
values of $\mugegm$ of 0.02 at $Q^2\approx 1$~(GeV/c)$^2$ and 0.05 for $Q^2
\approx 5$~(GeV/c)$^2$ would yield a modified TPE correction, changing the
extracted value of $a(Q^2)$ by 0.003--0.005 (and the low $\varepsilon$ value
of $R_{e^{+} e^{-}}$ in Fig.~\ref{fig:TPEcompare} by 0.01--0.02), which is
typically comparable to the total uncertainty we quote in the fit
Eq.~(\ref{eq:globfit}) and even larger for the lower $Q^2$ values of our fit.
This shift in $\mugegm$ is at the level of the uncertainties in the
polarization extraction.  It is also comparable to the level at which the
$\varepsilon$ dependence of the TPE contributions to the polarization data are
constrained.  Thus, the uncertainty neglected when the uncertainties and/or
TPE contributions to the PT data are neglected are comparable to the total
uncertainties obtained in these extractions.

There has been a recent push to make new measurements of the ratio $R_{e^{+}
e^{-}}$, focusing on small $\varepsilon$, where TPE contributions appear to be
largest.  The first is the VEPP-3 experiment~\cite{vepp_proposal,
nikolenko2010a, nikolenko2010b}, where the internal target at the VEPP-3
electron-positron storage ring at Novosibirsk was used to extract the ratio
$R_{e^{+} e^{-}}$  at $Q^2=1.60$~(GeV/c)$^2$ and $\varepsilon \approx 0.4$. A
raw $R_{e^{+} e^{-}}$ ratio of 1.056$\pm$0.011 was obtained, which must be
reduced by the charge-dependent bremsstrahlung correction, estimated to be
$\sim$3\%.  The Blunden~\etal calculation~\cite{blunden05a} predicts
$R_{e^{+} e^{-}}=1.036$ (1.043 with the additional contribution included in
Ref.~\cite{arrington07}). Our global fit predicts $R_{e^{+} e^{-}} =
1.060\pm0.009$, but the $Q^2$ value is low enough that the extraction is not
expected to be very reliable, and the uncertainty associated with the
model dependence of the extraction is larger than the quoted fit uncertainty.

The second experiment is Jefferson Lab experiment E07-005~\cite{e07-005},
where a mixed beam of $e^{+}$ and $e^{-}$ produced via pair production from a
secondary photon beam, was used to simultaneously measure
$\sigma(e^{+}p$$\rightarrow$$e^{+}p)$ and
$\sigma(e^{-}p$$\rightarrow$$e^{-}p)$ elastic scattering cross sections. Cross
sections in the kinematical range of $0.5 < Q^2 < 2.0$ (GeV/c)$^2$ and
$0.2 < \varepsilon < 0.9$ can be measured. The third is the OLYMPUS
experiment~\cite{kohl09}, where the DORIS lepton storage ring at DESY will be
used to extract the ratio $R_{e^{+} e^{-}}$ from $Q^2=$ 0.6 (GeV/c)$^2$ and
$\varepsilon=$ 0.90, to $Q^2=$ 2.2 (GeV/c)$^2$ and $\varepsilon=$ 0.35.

\section{conclusions} \label{conclusions}

In conclusion, we extracted the elastic electromagnetic form factors of the
proton using two different parametrizations for the OPE-TPE interference
function $F(Q^2,\varepsilon)$: the QA parametrization
Eq.~(\ref{eq:qattanfit2}) and the BK parametrization
Eq.~(\ref{eq:Kobushkin3}). Both parametrizations are linear in $\varepsilon$,
but make different assumptions as to where the TPE contributions vanish. In
the BK parametrization, the TPE correction to $\sigma_{R}$ was constrained by
enforcing the Regge limit which was not done in the QA parametrization. In
both parametrizations, we constrained $\mu_{p}G_{Ep}/G_{Mp}$ using recoil
polarization data.  The values of $G_{Mp}$ and $G_{Ep}$ extracted using the QA
parametrization are smaller than those obtained using the BK parametrization,
by as much as 10\%, with significantly larger uncertainties obtained using the
QA parametrization.

The form factor results from the BK fit are generally in good agreement with
the form factors based on a global analysis including calculated TPE
corrections~\cite{arrington07}, as well as some previous phenomenological
extractions.  For $\gep$, this is essentially because the polarization
transfer data is assumed to be unaffected by TPE corrections and taken as a
constraint in most of the extractions, while for $\gmp$, the extraction is
slightly more dependent on the detailed assumptions of the analyses. Using the
BK parametrization, the TPE amplitude $a(Q^2)$ was extracted. The amplitude is
on the few percent level, and increases in size with increasing $Q^2$. We 
parametrize the TPE contribution and fit uncertainties as $a(Q^2) =
(-0.0191\pm0.0014)\sqrt{Q^2} \pm 0.003$ for $1.5 < Q^2 < 4.0$~(GeV/c)$^2$.
We estimate that the uncertainties associated with the assumptions made
in this extraction (no TPE correction to polarization data, linear of TPE
contribution to the cross section, etc...), common to most of the extractions
presented here, yield an uncertainty that is comparable to the quoted
fit uncertainty, although these uncertainties are more likely to be strongly
correlated with $Q^2$ and of special importance at lower $Q^2$ values.  Note
that recent high-precision polarization measurements of $\mugegm$ at low
$Q^2$~\cite{bernauer10, e05017, zhan11, ron11}, combined with expected results
from comparison of positron and electron scattering, will significantly
improve our knowledge of the TPE contributions at lower $Q^2$ values. 
However, at very low $Q^2$ values, the TPE contributions have significantly
more impact on $\gmp$~\cite{arrington11c, bernauer11}, which is more difficult
to extract precisely in Rosenbluth measurements.

Note that the cross sections reported in Ref.~\cite{qattan05} were determined
by detecting recoiling protons, in contrast to all other measurements which
detected the scattered electrons. The consistency of the extraction suggests
that the approximations used to calculate standard radiative corrections,
which yield very different corrections for these two cases, are reliable,
although the quantitative comparison is limited by the precision of the
electron detection measurements.  A more extensive set of such measurements,
covering $0.4 < Q^2 \ltorder 5$~(GeV/c)$^2$ is under analysis, and will allow for
a much more detailed examination~\cite{e05017}.

Finally, we compare extractions of the TPE contributions to the ratio of
positron--proton and electron--proton scattering cross sections.  We use these
to make predictions for the higher $Q^2$ kinematics of the recently completed
and ongoing measurements of $R_{e^{+} e^{-}}$.  The lower $Q^2$ values enter
the region where the present data are limited in their ability to extract TPE
contributions based on the comparison of cross section and polarization
measurements.  The positron measurements will provide new information which
can provide the first direct experimental evidence for TPE contributions at
low $Q^2$.

\begin{acknowledgments}

This work was supported by Khalifa University of Science, Technology and
Research and by the U.~S. Department of Energy, Office of Nuclear Physics,
under contract DE-AC02-06CH11357. We thank Mrs. Phyllis Burns and Dr. Nicolas
Moore for reading the manuscript and making valuable comments and suggestions.
We also thank the IT department at Khalifa University for their technical
assistance.

\end{acknowledgments}

\bibliography{longpaper_TPEparam}

\end{document}